\documentclass[twocolumn,10pt]{tsfp}
\usepackage{flushend}
\usepackage{graphicx}
\usepackage[authoryear,round]{natbib}
\usepackage{fancyhdr}
\usepackage{amsmath}
\usepackage{amssymb}
\DeclareMathAlphabet{\mathcal}{OMS}{cmsy}{m}{n}

\usepackage[normalem]{ulem}
\usepackage{color}

\usepackage[ruled, vlined]{algorithm2e}

\title{On the use of recurrent neural networks for predictions of turbulent flows}

\author{Luca Guastoni
    \affiliation{
	Linn\'e FLOW Centre, KTH Mechanics, \\
	Swedish e-Science Research Centre (SeRC)\\
	SE-100 44 Stockholm, Sweden\\
    guastoni@mech.kth.se
    }	
}

\author{Prem A. Srinivasan
    \affiliation{
	Linn\'e FLOW Centre, KTH Mechanics, \\
	Swedish e-Science Research Centre (SeRC)\\
	SE-100 44 Stockholm, Sweden\\
    paas2@kth.se
    }	
}

\author{Hossein Azizpour
    \affiliation{
	School of Elect. Eng. and Computer Science, KTH \\
	Swedish e-Science Research Centre (SeRC)\\
	SE-100 44 Stockholm, Sweden\\
    azizpour@kth.se
    }	
}

\author{Philipp Schlatter
    \affiliation{
	Linn\'e FLOW Centre, KTH Mechanics, \\
	Swedish e-Science Research Centre (SeRC)\\
	SE-100 44 Stockholm, Sweden\\
    pschlatt@mech.kth.se
    }	
}

\author{Ricardo Vinuesa
    \affiliation{
	Linn\'e FLOW Centre, KTH Mechanics, \\
	Swedish e-Science Research Centre (SeRC)\\
	SE-100 44 Stockholm, Sweden\\
    rvinuesa@mech.kth.se
    }	
}

\begin{document}

\maketitle   
\thispagestyle{fancy}

\fontsize{9}{11}\selectfont

\section*{ABSTRACT}

In this paper, the prediction capabilities of recurrent neural networks are assessed in the low-order model of near-wall turbulence by Moehlis {\it et al.} (New J. Phys. {\bf 6}, 56, 2004). Our results show that it is possible to obtain excellent predictions of the turbulence statistics and the dynamic behavior of the flow with properly trained long short-term memory (LSTM) networks, leading to relative errors in the mean and the fluctuations below $1\%$. We also observe that using a loss function based only on the instantaneous predictions of the flow may not lead to the best predictions in terms of turbulence statistics, and it is necessary to define a stopping criterion based on the computed statistics. Furthermore, more sophisticated loss functions, including not only the instantaneous predictions but also the averaged behavior of the flow, may lead to much faster neural network training.


\section*{INTRODUCTION}

The use of neural networks (NNs) in the context of turbulent flows has recently started to receive increasing attention, as discussed for instance by \cite{duraisamy_et_al}. Neural networks are computational frameworks used to learn certain tasks from examples, and they are a tool widely used in machine learning. Their success in a number of areas, mainly related to pattern recognition, can be attributed to the increase in available computational power (mainly through graphics processing units, {\it i.e.} GPUs) and it explains the increasing interest in their use for turbulence \citep{kutz}. Several studies have explored the possibility of using neural networks to develop more accurate Reynolds-averaged Navier--Stokes (RANS) models \citep{wu_et_al}, while other studies  aim at developing subgrid-scale (SGS) models for large-eddy simulations (LESs) of turbulent flows \citep{lapeyre_et_al}. Other relevant applications include the development of robust inflow conditions for high-Reynolds-number turbulence simulations \citep{fukami_et_al} and the identification of coherent structures in the flow \citep{jimenez_ml}.

The aims of the present work are to assess whether it is possible to use NNs to predict the temporal dynamics of turbulent shear flows, and to test various strategies to improve such predictions. In order to easily obtain sufficient data for training and validation, we considered a low-order representation of near-wall turbulence, described by the model proposed by \cite{moehlis_et_al}. The mean profile, streamwise vortices, the streaks and their instabilities as well as their coupling are represented by nine spatial modes $\mathbf{u}_{j}(\mathbf{x})$. The spatial coordinates are denoted by $\mathbf{x}$ and $t$ represents time. The instantaneous velocity fields can be obtained by superimposing the nine modes as: $\mathbf{\tilde{u}}(\mathbf{x},t) = \sum_{j=1}^9 a_{j}(t) \mathbf{u}_j(\mathbf{x})$, where Galerkin projection can be used to obtain a system of nine ordinary differential equations (ODEs) for the nine mode amplitudes $a_{j}(t)$. A model Reyonlds number $Re$ can be defined in terms of the channel full height $2h$ and the laminar velocity $U_{0}$ at a distance of $h/2$ from the top wall. Here we consider $Re=400$ and employ $U_{0}$ and $h$ as velocity and length scales, respectively. The ODE model was used to produce over 10,000 time series of the nine amplitudes, each with a time span of 4,000 time units, for training and validation. The domain size is $L_{x}=4 \pi$, $L_{y}=2$ and $L_{z}=2 \pi$, where $x$, $y$ and $z$ are the streamwise, wall-normal and spanwise coordinates, and we consider only time series that are turbulent over the whole time span. In the next sections we will discuss the feasibility of using various types of neural network to predict the temporal dynamics of this simplified turbulent flow. All the results discussed in this study were obtained using the machine learning software framework developed by Google Research called TensorFlow \citep{tensor_flow}.

\section*{PREDICTIONS WITH RECURRENT NEURAL NETWORKS}
The simplest type of neural network is the so-called multilayer perceptron (MLP) \citep{rumelhart1985learning}, which consists of two or more layers of nodes (also denoted by the term neurons), where each node is connected to the ones in the preceding and succeeding layers. Although MLPs are frequently used in practice, their major limitation is that they are designed for point prediction as opposed to time-series prediction, which might require a context-aware method. Nevertheless, MLPs provide a solid baseline in machine-learning applications and thereby help verifying the need for a more sophisticated network architecture. We first assessed the accuracy of MLP predictions of the nine-equation model by \cite{moehlis_et_al}, where the time evolution of the nine coefficients was predicted with several different architectures. The turbulence statistics were obtained by averaging over the periodic directions ({\it i.e.}\ $x$ and $z$) and in time over 500 complete time series, which was sufficient to ensure statistical convergence in this case \citep{srinivasan}. In order to quantify the accuracy of the predictions, we consider the relative error between the model and the MLP prediction (denoted by the subindices `mod' and `pred', respectively) for the mean flow as:
\begin{equation}
E_{\overline{u}}=\frac{1}{2\  {\rm max}(\overline{u}_{{\rm mod}})} \int_{-1}^{1} \left | \overline{u}_{{\rm mod}}-\overline{u}_{{\rm pred}} \right |  {\rm d}y,
\end{equation}
where the normalization with the maximum of $\overline{u}$ is introduced to avoid spurious error estimates close to the centerline where the velocity is 0. This error is defined analogously for the streamwise velocity fluctuations $\overline{u^{2}}$. A number of MLP architectures were investigated \citep[see additional details in the work by][]{srinivasan}, and the best predictions were obtained when considering $l=5$, $n=90$ and $p=500$, which denote respectively the number of hidden layers, the number of neurons per layer and the number of previous $a_{j}(t)$ values used to obtain a prediction. With this architecture, the errors in the mean and fluctuations are $E_{\overline{u}}=3.21\%$ and $E_{\overline{u^{2}}}=18.61\%$ respectively, indicating that although acceptable predictions of the mean flow can be obtained, the errors in the fluctuations are high. Furthermore, the size of the input is $d=9p=4,500$ ({\it i.e.} 9 coefficients over the past 500 time steps are used to predict the next 9 coefficients), which is quite large. Since the MLP performs point predictions, it does not exploit the sequential nature of the data, and it is therefore important to assess the feasibility of using other types of networks, {\it i.e.} the so-called recurrent neural networks (RNNs), which can benefit from the information contained by the temporal dynamics in the data.

\begin{figure*}
    \centering
	\includegraphics[width=4.5in]{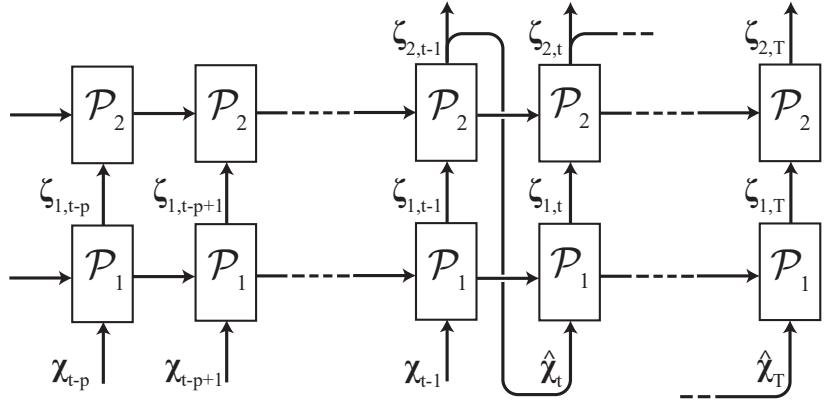}
	\caption{``Unrolled" representation of a multi-layer LSTM, where $\mathcal{P}_i$ is the set of parameters that characterize the LSTM unit of the \textit{i}-th layer. Note that $\mathcal{P}_i$ is shared among all the $p$ time steps considered for the prediction. Here $\pmb{\chi}$ is an input based on the nine-equation model, whereas $\pmb{\hat{\chi}}$ is predicted by the neural network and  $T$ is the final time step of the prediction.}
	\label{figure2}
\end{figure*}

In its simplest form, an RNN is a neural network containing a single hidden layer with a feedback loop. As opposed to MLPs, each node of the RNN layer has an internal state vector that is combined with the input vector to compute the output. The output of the hidden layer in the previous time instance is fed back into the hidden layer along with the current input. This allows information to persist, making the network capable of learning sequential dependencies. In practice, this simple recurrent network is not effective to learn long-term dependencies, hence a more sophisticated model is required, such the long short-term memory (LSTM) network proposed by \cite{hochreiter_schmidhuber}, or the gated recurrent unit (GRU) network developed by \cite{cho}. Both architectures use a gating mechanism to actively control the dynamics of the recurrent connections. Each unit in the LSTM layer performs four operations through three different gates. The \textit{forget gate} uses the output in the previous time instance $\pmb{\zeta}_{t-1}$ and the current input $\pmb{\chi}_{t}$ to determine which part of the cell state $\mathbf{C}_{t-1}$ should be retained in the current evaluation. The \textit{input gate} uses the same quantities to determine which values of the cell state should be updated and it also computes the candidate values for the update. Finally the \textit{output gate} uses the newly-updated cell state to compute the output. Algorithm~\ref{algo_rnn} illustrates how the output is computed and how the cell state is updated, where $\otimes$ indicates the Hadamard product and $\sigma$ denotes the logistic sigmoid function. A schematic representation of a multi-layer LSTM is shown in Figure~\ref{figure2}. The model is defined by a set of parameters $\mathcal{P}$ which comprise the weight matrices $\mathbf{W}$ and the biases $\mathbf{b}$. During training, the values of the parameters are optimized to minimize a certain loss function.
\begin{algorithm}[t!]
\DontPrintSemicolon
\KwIn{Sequence $\pmb{\chi}_{1}, \pmb{\chi}_{2}, \dots \pmb{\chi}_{p}$}
\KwOut{Sequence $\pmb{\zeta}_{1}, \pmb{\zeta}_{2}, \dots \pmb{\zeta}_{p}$}
 set $\mathbf{h}_0 \leftarrow 0$\;
 set $\mathbf{C}_0 \leftarrow 0$\;
 \For{$t\leftarrow 1$ \KwTo $p$}{
  \ \ $\mathbf{f}_t \leftarrow \sigma(\mathbf{W}_{f} [\pmb{\chi}_{t}, \pmb{\zeta}_{t-1}] + \mathbf{b}_{f})$\;
  \ \ $\mathbf{i}_t \leftarrow \sigma(\mathbf{W}_{i} [\pmb{\chi}_{t}, \pmb{\zeta}_{t-1}] + \mathbf{b}_{i})$\;
  \ \ $\mathbf{\widetilde{C}}_t \leftarrow \tanh(\mathbf{W}_{f} [\pmb{\chi}_{t}, \pmb{\zeta}_{t-1}] + \mathbf{b}_{f})$\;
  \ \ $\mathbf{C}_t \leftarrow \mathbf{f}_t\otimes\mathbf{C}_{t-1} + \mathbf{i}_t\otimes\mathbf{\widetilde{C}}_t$\;
  \ \ $\mathbf{o}_t \leftarrow \sigma(\mathbf{W}_{o} [\pmb{\chi}_{t}, \pmb{\zeta}_{t-1}] + \mathbf{b}_{o})$\;
  \ \ $\pmb{\zeta}_t \leftarrow \mathbf{o}_t\otimes\tanh(\mathbf{C}_{t-1})$\;
 }
 \caption{Compute the output sequence of an LSTM network.}
 \label{algo_rnn}
\end{algorithm}

\begin{figure*}
\centering
\vspace{0pt}
\includegraphics[width=0.82 \textwidth]{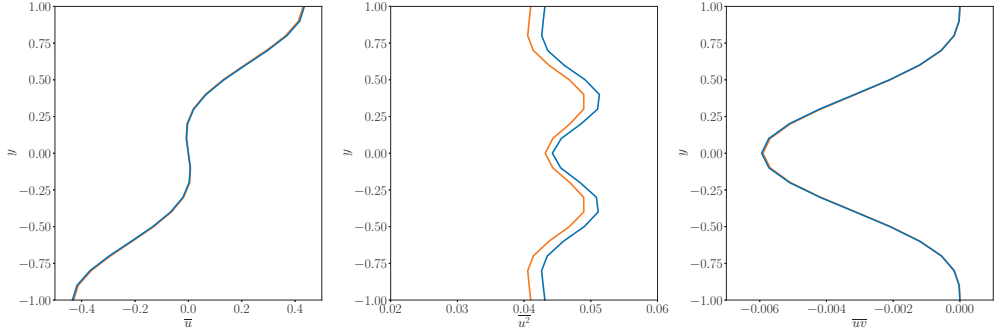}
\caption{Turbulence statistics corresponding to (left) streamwise mean profile, (middle) streamwise velocity fluctuations and (right) Reynolds shear stress. Orange is used for the reference nine-equation model and blue for the predictions using an LSTM network with 1 layer and 90 neurons, trained with 10,000 datasets.}
\label{fig_stats}
\end{figure*}

We initially analyzed the prediction capabilities of LSTM networks for this turbulent shear flow wall model by considering a network with a single layer of 90 LSTM units. We trained it with three different datasets, consisting respectively of 100, 1,000, and 10,000 time series spanning 4,000 time units each \citep{srinivasan}. We considered a validation loss defined as the  sum  over $p$ time steps of the squared error in the prediction of the instantaneous coefficients $a_{j}$, and observed that better predictions of the turbulence statistics could be obtained when larger datasets were employed for training. Note that we considered $20\%$ of the training data as a validation set, which is then used to test the evolution of the validation loss on data which has not been seen by the network during training. Using 10,000 time series for training, we obtained excellent predictions of the turbulence statistics, with $E_{\overline{u}}=0.45\%$ and $E_{\overline{u^{2}}}=2.49\%$. This was obtained with $p=10$, {\it i.e.} with an input size 50 times smaller than that used with MLP. In Figure~\ref{fig_stats} we show a comparison of the turbulence statistics obtained from the nine-equation model and this LSTM network, including the mean flow, the fluctuations and the Reynolds shear-stress profile $\overline{uv}$. The agreement of all the statistics with the reference data is excellent, and even higher-order moments exhibit low relative errors, {\it i.e.}  $1.01\%$ and $2.57\%$ for skewness and flatness, respectively \citep{srinivasan}. These results highlight the excellent predicting capabilities of the LSTM network, given that sufficient training data is provided, due to the ability of the network to exploit the sequential nature of the data.

\begin{figure}[t]
\centering
\includegraphics[width=2.4in]{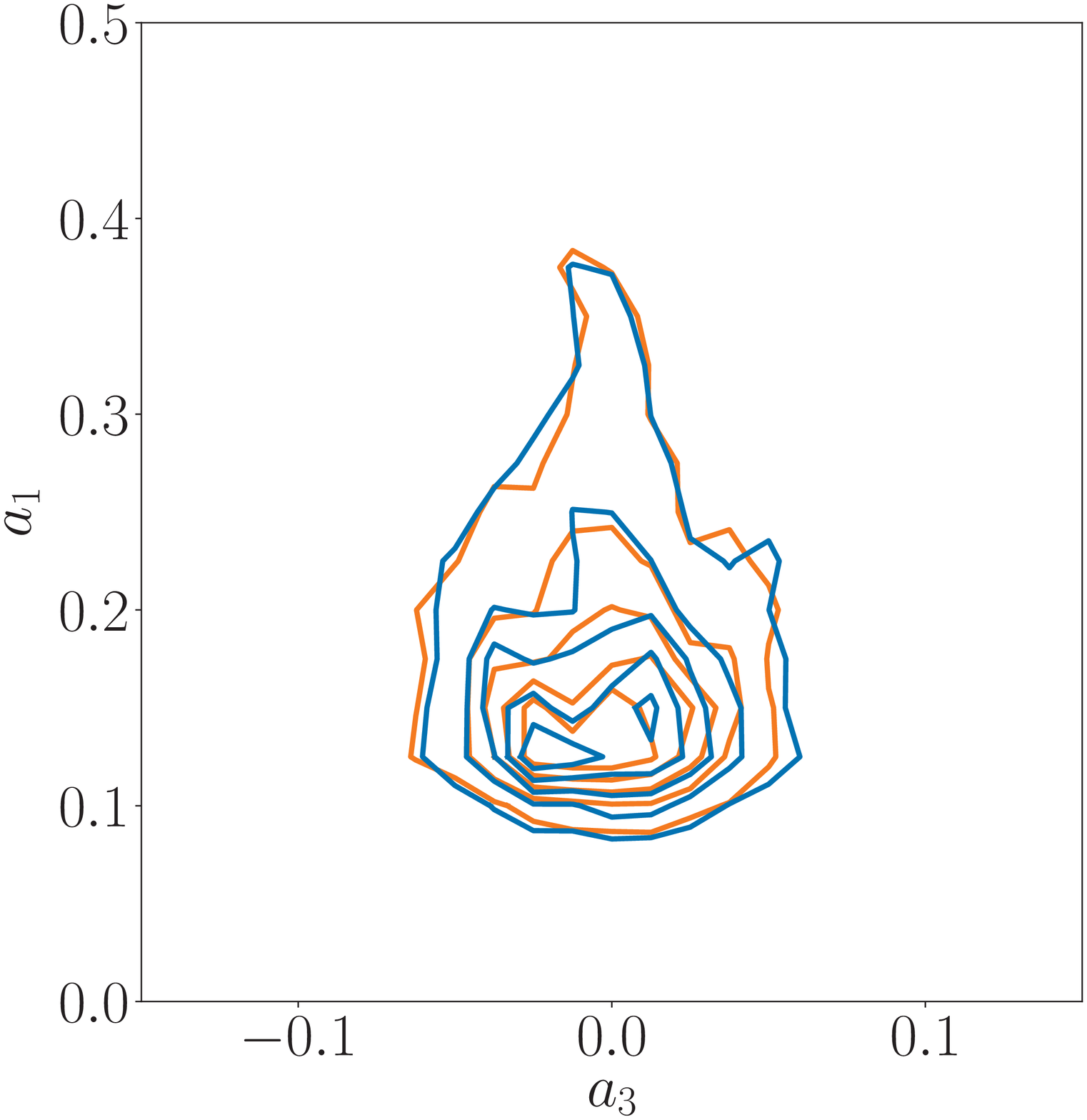}
\includegraphics[width=2.4in]{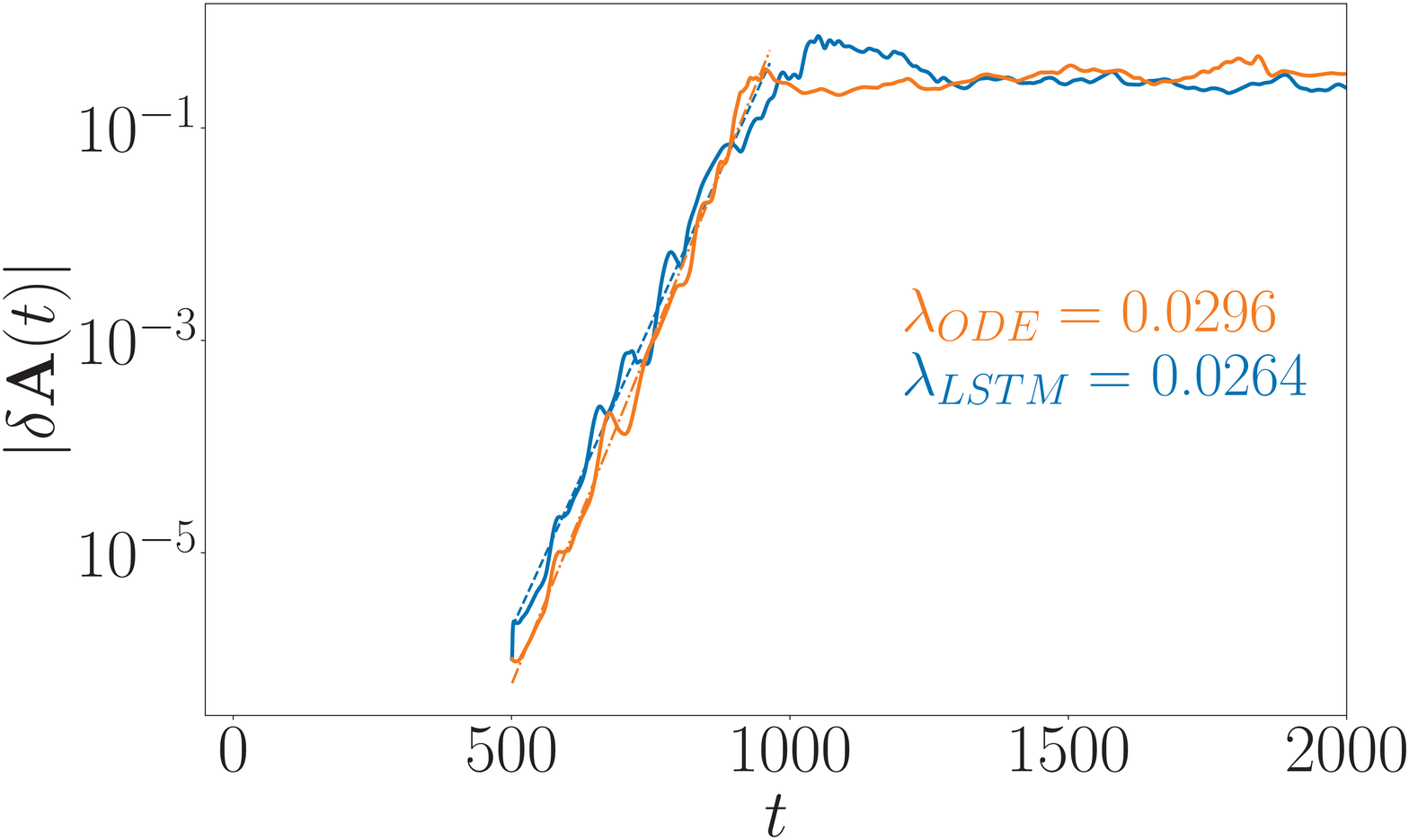}
\caption{(Top)  Probability density function of the Poincar\'e maps, where the intersection with the $a_{2}=0$ plane (with ${\rm d}a_{2}/{\rm d}t <0$) is shown. (Bottom) Ensemble-averaged divergence of instantaneous time series after a perturbation with $| \delta \mathbf{A}_{0} |=10^{-6}$ is introduced at $t_{0}=500$, showing initial exponential growth and the value of the Lyapunov exponent (dashed lines added to illustrate the obtained slope). In both panels orange and blue denote reference model and LSTM prediction, respectively.}
\label{fig_dynamics}
\end{figure}

The quality of the predictions was further assessed in terms of the dynamic behavior of the system, first through the Poincar\'e map defined as the intersection of the flow state with the hyperplane $a_{2}=0$ on the $a_{1}-a_{3}$ space (subjected to ${\rm d}a_{2}/{\rm d}t <0$). This map essentially shows the correlation between the amplitudes of the first and third modes, {\it i.e.} the modes representing the laminar profile and the streamwise vortices in the nine-equation model. In Figure~\ref{fig_dynamics}~(top) we show the probability density function (pdf) of the Poincar\'e maps constructed from the 500 time series obtained from the LSTM prediction and the reference nine-equation model. In this figure it can be observed that the LSTM network captures the correlation between the amplitudes of both modes, which indicates that their interaction is adequately represented by the NN. We also studied the separation among trajectories in the reference model and in the LSTM predcition by means of Lyapunov exponents. For two time series 1 and 2, we define the separation of these trajectories as the Euclidean norm in nine-dimensional space: 
 \begin{equation}
\left | \delta \mathbf{A}(t) \right | = \left [ \sum_{i=1}^{9} \left (a_{i,1}(t)-a_{i,2}(t)  \right )^{2} \right ]^{1/2},
\end{equation} 
and denote the separation at $t=t_{0}$ as $ | \delta \mathbf{A}_{0} |$. The initial divergence of both trajectories can be assumed to behave as: $\left | \delta \mathbf{A}(t') \right | = \exp (\lambda t') \left | \delta \mathbf{A}_{0} \right |$, where $\lambda$ is the so-called Lyapunov exponent and $t'=t-t_{0}$. We introduced a perturbation with norm $| \delta \mathbf{A}_{0} | =10^{-6}$ (which approximately corresponds to the accuracy of the current LSTM architecture) at $t_{0}=500$, where all the coefficients are perturbed, and analyzed its divergence with respect to the unperturbed trajectory. In Figure~\ref{fig_dynamics}~(bottom) we show the evolution of $ | \delta \mathbf{A}(t) |$ with time for the reference and the LSTM prediction, after ensemble averaging 10 time series. Both rates of divergence are very similar, with almost identical estimations of the Lyapunov exponents $\lambda$: 0.0264 for the LSTM and 0.0296 for the nine-equation model. Also note that after around approximately 1,000 time units of divergence, both curves saturate. This result provides additional evidence supporting the excellent predictions of the dynamic behavior of the original system when using the present LSTM architecture.




\section*{TOWARDS IMPROVING NEURAL NETWORK PREDICTIONS}

We have shown the potential of a particular type of RNN, the LSTM, to accurately predict the temporal dynamics and statistics of a low-dimensional representation of near-wall turbulence. Next we explore different strategies to potentially improve the accuracy and efficiency of RNN predictions.

\subsection*{Validation loss and training stopping criterion}

\begin{figure}[t]
    \centering
    \includegraphics[width=3in]{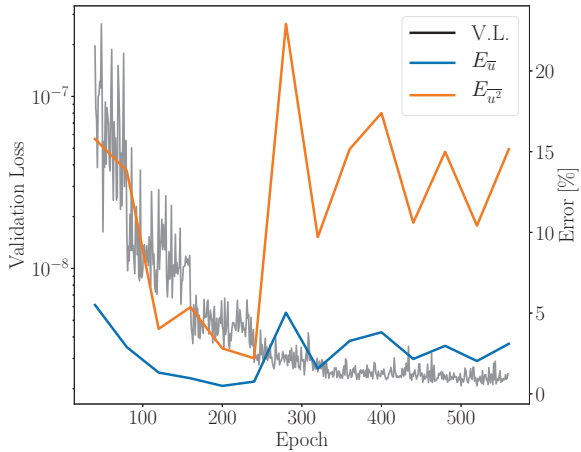}
    \caption{Evolution of the validation loss and the statistical errors as the training of the LSTM2--1--100 network progresses.}
    \label{err_vs_val}
\end{figure}

As discussed above, the amplitudes of the modes in the model by \cite{moehlis_et_al} exhibit fluctuations that are compatible with a chaotic turbulent state. Given the high sensitivity of the model to very small variations in the mode amplitudes, a loss function based on short-time horizon predictions, namely one time step ahead, is required to obtain satisfactory predictions. On the other hand 
the trained model needs to correctly reproduce not only the instantaneous behavior but also the statistical features of the original shear flow model. The approach used in the work by \cite{srinivasan} involves a loss function based only on the error in the instantaneous prediction. Neural networks having at least one hidden layer have been shown to be universal approximators \citep{Cybenko}, hence they are in principle able to represent any real function. A perfect reproduction of the temporal behavior of the model would also provide correct mean and turbulent fluctuations at no added cost, however there is no guarantee that such a model can be learned and, even in that case, the model would theoretically be available after an infinitely long training. To verify to which extent the loss function based on instantaneous prediction represents an effective solution, different neural network configurations were tested to assess the correlation between the achieved validation loss and the error in the statistics of the flow. In Table 1 we summarize the various LSTM architectures under study, where we vary the number of layers, the number of time series used for training and the time step between samples. 
Let us consider the case LSTM2--1--100, consisting of 2 layers, with 90 units per layer, trained with 100 time series and a timestep of 1. Figure~\ref{err_vs_val} shows the validation loss and the relative errors $E_{\overline{u}}$ and $E_{\overline{u^{2}}}$ for this network, as functions of the number of epochs trained ({\it i.e.}\ the number of complete passes through all the samples contained in the training set). In the initial stage of the training, starting from the randomized initialization of the weights and biases, the reduction of the error in the instantaneous behavior and in the statistical quantities show a similar trend. However, this figure also shows that lower validation loss values do not always lead to a better approximation of the turbulence statistics. In fact, as the training progresses, the optimization algorithm continues to improve the short-term predictions, whereas beyond around 240 epochs the error in the statistics does not follow a descending trend anymore. The observed behavior is explained by the fact that the loss function does not contain any term explicitly related to the statistics which could guide the optimization algorithm towards parameter sets with a better representation of the statistical quantities. Note that since the initialization of the parameters of the network is random, the performance in the prediction of mean and fluctuation may vary when the same model is trained multiple times. The achievable accuracy and the epoch at which this value will be reached are unknown \textit{a priori}.

These results indicate that different strategies can be implemented in order to reduce the error on the statistics of the flow. One possible approach consists in including a new term in the loss function accounting for the error in the turbulence statistics. In this case the relative importance of the two terms needs to be adjusted, as prioritizing the accuracy of the statistics may lead to a model that learns only the average behavior of the system. Alternatively, it is possible to use the fact that the time horizon of the predictions influences which features of the problem are learnt by the neural network, as highlighted by \cite{Chiappa}. In that work it was shown how improvements in the short-term accuracy ({\it i.e.}\ in the prediction of the instantaneous behavior) come at the expense of the accuracy of global dynamics. Using the results of the network to make predictions several time steps ahead would encourage the network to learn the long-term behavior of the system and thus its global dynamics. As stated by \cite{Chiappa}, this approach has the added advantage of training the model in a way that is similar to its actual utilization. In fact, during the evaluation and usage, our networks rely only on the previous predictions after the first $p$ time steps. Note however that taking into account the error in the current prediction based on previous predicted values typically results in a much more complex loss function. Both approaches require additional hyper-parameters that need to be optimized in order to obtain a satisfactory performance. In this study we aim at keeping a simple loss function, and we use the error in the statistics as criterion to halt the training when a minimum is reached for this value. Note that the error can vary significantly from one epoch to the other, hence it is advisable to consider multiple epochs to identify the general trend of the error curves. Doing so, we can achieve excellent predictions of the turbulence statistics while using a simple loss function based on the instantaneous predictions of the coefficients. As shown in Table~\ref{table_lstm}, the improvement over the models reported in our previous work \citep{srinivasan} is particularly evident for the models trained on the small dataset, yielding an accuracy in the statistics comparable with that of the networks trained with bigger data sets. It is also important to note that the improved scheduled reduction of the learning rate employed for the results in Table~\ref{table_lstm} allowed to obtain much lower validation losses than in our previous work, using a similar training time. When reaching such low values of the loss function, the trade-off between the instantaneous and the average performance is more apparent.

\subsection*{Effect of the time step}
The sequences provided to the neural network for training are evenly spaced in time, however the choice of the proper time step between data points $\Delta t$ depends on the problem at hand. The time step acts as a low-pass filter on the data, preventing the model from learning higher-frequency dynamics. On the other hand, for a fixed amount of samples, a larger $\Delta t$ allows to train the model using more time series. As shown in Table~\ref{table_lstm}, we considered the LSTM network with 1 layer and 90 neurons, and trained it using the same time series with $\Delta t=10$, 1 and 0.1 time units. Note that the input dimension is maintained constant by setting $p=10$. The number of time series and epochs for training were chosen so that it could be possible to compare models that have been trained on a similar number of samples. The results in Table~\ref{table_lstm} show that increasing the time step from 1 to 10 leads to a validation loss three orders of magnitude larger, a fact that indicates the difficulty in learning the model dynamics when such a coarse sampling in time is considered. On the other hand, reducing the time step from 1 to 0.1 does not yield any additional improvement in the predictions. The loss function has a similar trend and the final values are comparable when using time steps equal to 1 and 0.1, showing that most of the characteristics of the system have been properly captured. It may be possible to find a $\Delta t$ that further reduces the error based on the temporal characteristics of the signal. 
\begin{table*}[ht]
 \centering
 \caption{Summary of LSTM cases and their performance using different numbers of training data sets and time resolutions. Note that we employed 90 units and $p=10$ in all the cases. The statistical errors for LSTM1--10--1000 are not reported because the predictions exhibited a clearly non-physical behavior during all stages of training.}
 \label{table_lstm}
 \begin{tabular}{c c c c c c c}
 & & \\  
 \hline
 \hline
 Case & N. Layers & $\Delta t$ & Training data sets & $E_{\overline{u}}$ $[\%]$ & $E_{\overline{u^{2}}}$ $[\%]$ & Validation Loss\\ 
 \hline
LSTM1--1--100 & 1 & 1 & 100 & 0.26 & 0.59 & $ 6.68 \times 10^{-9}$ \\
LSTM1--01--100 & 1 & 0.1 & 100 & 1.81 & 6.03 & $ 9.13 \times 10^{-10}$ \\
LSTM1--10--1000 & 1 & 10 & 1,000 & -- & -- &  $ 3.65 \times 10^{-5}$ \\
LSTM1--1--1000 & 1 & 1 & 1,000 & 0.57 & 0.58 & $ 8.36 \times 10^{-9}$ \\
LSTM1--01--1000 & 1 & 0.1 & 1,000 & 1.18 & 1.39 & $ 6.46 \times 10^{-9}$ \\
LSTM1--1--10000 & 1 & 1 & 10,000 & 0.31 & 0.48 &  $ 9.85 \times 10^{-9}$ \\
LSTM2--1--100 & 2 & 1 & 100 & 0.80 & 1.13 & $ 8.39 \times 10^{-9}$ \\
LSTM2--1--1000 & 2 & 1 & 1,000 & 0.54 & 0.62 & $ 8.84 \times 10^{-9}$ \\
LSTM2--1--10000 & 2 & 1 & 1,000 & 0.69 & 1.37 & $ 2.72 \times 10^{-9}$ \\
 \hline
 \hline
 \end{tabular}
 \end{table*}

\subsection*{Use of gated recurrent units (GRUs)}
The performance of an alternative type of RNN, the so-called gated recurrent unit (GRU), is also studied here. The structure of GRU layers is simpler than in the LSTM, consisting of a single \textit{update gate} instead of the forget and input gates. Also, the cell state and the output are merged into a single vector. The network architecture considered here has 1 layer of 90 nodes and it is similar in every aspect to the corresponding LSTM case, except for the node definition. The number of parameters that need to be optimized is smaller than in the LSTM, and therefore GRUs should require less computational resources to be trained. In our experience however, when training the considered architecture on CPU, the LSTM network was approximately as fast as its GRU counterpart. Despite the fact that it is possible to obtain similar validation losses with GRUs and LSTM networks, the resulting errors in the statistics are significantly higher in the former. In particular, when training with only 100 time series the predicted results exhibited a non-physical behavior. Although the results in Table~\ref{table_gru} suggest that the predictions may improve when using much larger training databases, the LSTM networks provide much more accurate predictions and they are therefore preferred for the present application.

\begin{table*}[ht]
 \centering
 \caption{Summary of GRU cases and their performance using different numbers of training data sets. Note that in all the cases 1 layer of 90 units was employed, with $p=10$. The statistical errors for GRU100 are not reported because the predictions exhibited a clearly non-physical behavior during all stages of training.}
 \label{table_gru}
 \begin{tabular}{c c c c c}
 & & \\  
 \hline
 \hline
 Case & Training data sets & $E_{\overline{u}}$ $[\%]$ & $E_{\overline{u^{2}}}$ $[\%]$ & Validation Loss\\ 
 \hline
GRU100 & 100 & -- & -- & $1.33 \times 10^{-8}$ \\
GRU1000 & 1,000 & 2.30 & 12.49 & $6.13 \times 10^{-9}$ \\
GRU10000 & 10,000 & 3.05 & 2.61 & $5.61 \times 10^{-9}$ \\
 \hline
 \hline
 \end{tabular}
 \end{table*}

\section*{SUMMARY AND CONCLUSIONS}

In this study we assessed the feasibility of using RNNs to predict the temporal dynamics of the low-order model of near-wall turbulence by \cite{moehlis_et_al}. Our previous results \citep{srinivasan} indicated that it is possible to obtain excellent predictions of the turbulence statistics using LSTM networks, and to reproduce the temporal dynamics of the system characterized through \emph{e.g.}\ Poincar\'e maps and Lyapunov exponents. Here we show that, even using relatively small LSTM networks trained with low numbers of time series, {\it e.g.}\ the LSTM1--1--100 case, it is possible to obtain very low errors in the mean and the fluctuations, {\it i.e.}\ $E_{\overline{u}}=0.26\%$ and $E_{\overline{u^{2}}}=0.59\%$. It is important to highlight that a loss function based only on the instantaneous predictions of the mode amplitudes may not lead to the best predictions in terms of turbulence statistics, and it is necessary to define a stopping criterion based on the values of $E_{\overline{u}}$ and $E_{\overline{u^{2}}}$. Our results also suggest that using more sophisticated loss functions, including not only the instantaneous predictions but also the averaged behavior of the flow, may lead to much faster neural network training. It is however remarkable that using a simple loss function based on instantaneous values we also obtained very good predictions of Poincar\'e maps and Lyapunov exponents. We also assessed the impact of the time step, where the best network performance was obtained with $\Delta t=1$. Additionally, we compared the performance of LSTM networks and GRUs, and the former clearly provided much better predictions. The methods described in this work can be extended for their use in generation of inflow conditions for turbulence simulations and the development of off-wall boundary conditions for high-$Re$ simulations.

\section*{ACKNOWLEDGMENTS}
The authors acknowledge the funding provided by the Swedish e-Science Research Centre (SeRC) and the Knut and Alice Wallenberg (KAW) Foundation. Part of the analysis was performed on resources provided by the Swedish National Infrastructure for Computing (SNIC) at PDC and HPC2N.

\bibliographystyle{tsfp}
\bibliography{tsfp}

\end{document}